\documentclass[conference]{IEEEtran}

\usepackage{cite}
\usepackage{amsmath,amssymb,amsfonts}
\usepackage{algorithmic}
\usepackage{graphicx}
\usepackage{textcomp}
\usepackage{xcolor}
\def\BibTeX{{\rm B\kern-.05em{\sc i\kern-.025em b}\kern-.08em
    T\kern-.1667em\lower.7ex\hbox{E}\kern-.125emX}}

\def\thicchline{\noalign{\hrule height.8pt}}

\usepackage[hidelinks]{hyperref}

\begin{document}

\title{\huge Sensitivity of DC Network Representation for GIC Analysis}

\author{\IEEEauthorblockN{Aniruddh~Mishra}
\IEEEauthorblockA{\textit{Department of Electrical and Computer Engineering} \\
\textit{University of Texas at Austin}\\
Austin, TX \\
aniruddh@utexas.edu}
\and
\IEEEauthorblockN{Arthur~K.~Barnes, Jose~E.~Tabarez, Adam~Mate}
\IEEEauthorblockA{\textit{Information Systems and Modeling Group} \\
\textit{Los Alamos National Laboratory}\\
Los Alamos, NM \\
\{abarnes, jtabarez, amate\}@lanl.gov.}

\thanks{Approved for public release; distribution is unlimited. LA-UR-24-31515}
\thanks{979-8-3315-4112-5/25/\$31.00 ©2025 IEEE}
}

\IEEEoverridecommandlockouts

\maketitle


\begin{abstract}
Geomagnetic disturbances are a threat to the reliability and security of our national critical energy infrastructures. These events specifically result in geomagnetically induced currents, which can cause damage to transformers due to magnetic saturation.
In order to mitigate these effects, blocker devices must be placed in optimal locations. Finding this placement requires a dc representation of the ac transmission lines, which this paper discusses. Different decisions in this process, including the method of representing the blocking devices, result in significant variations to the power loss calculations. To analyze these effects, we conclude the paper by comparing the losses on a sample network with different modeling implementations.
\end{abstract}


\section{Introduction}
Geomagnetic disturbances (GMDs) adversely affect energy infrastructures; they result in geomagnetically induced currents (GICs) throughout the transmission network, which enter the electric grid at low resistance paths to ground, particularly at neutral wires of transformers.
Though GMDs can result in a range of contingencies in the grid, the primary impact we will be investigating is the effect on high-voltage power transformers. Since most transformers are rated for only a few additional amps while performing in a linear operating region, the addition of GICs into the network could result in extended power outages \cite{boteler_modeling_2016, hutchins_effect_2011, mate21-pmsgmd-cascade}.

Modeling GIC flow consists of two steps: (1) the generation of a dc network from a given ac model, as GICs are pseudo-dc phenomenon; and (2) utilizing a simulation and analysis software to find solutions for power flows and mitigate GMD impact. It is possible to model GICs with an EMTP-type simulation \cite{haddadi2020evaluation}, but it results in high computational complexity and an inability to implement optimization for mitigation contexts. The method of dc equivalent circuits \cite{boteler_development_2014} is more computationally efficient, is feasible for use in an optimization context \cite{zhu_blocking_2015}, and can be coupled to a positive sequence equivalent circuit with a transformer saturation model \cite{Walling1991,overbye2012integration}. A summary of this method for typical power systems components is provided in \cite{boteler_modeling_2016}.

GIC blockers are designed to mitigate dc current without affecting the ac network. The operating concepts are similar to that of a capacitor with the added requirement of not introducing any voltage transients, such as ferroresonance \cite{Kappenman1991}.

There are several techniques to optimize placements with a given blocker type. They range from extensively searching every combination of open and closed lines in the dc network to implementing  a mixed-integer optimization problem \cite{zhu_blocking_2015, barnes2024review}. Overall, the problem can be stated as finding the minimum total power loss with the number of blockers, $B$, less than or equal to a maximum quantity, $N_B$. Previous work has studied the sensitivity of GIC flow with respect to network parameters \cite{zheng_effects_2014}, specifically the impact of line resistance and grounding resistance on GIC magnitudes. However, the sensitivity of different dc network representations on the blocker placement problem has not been analyzed. Additionally, the partial blocker placement problem is a challenging and active area of research as placing blockers at one substation can increase GIC-induced transformer reactive power loss ($Q_{loss}$) at other substations \cite{zhu_blocking_2015}. A better understanding of the trade-offs between different blocker representations will help researchers develop scalable algorithms for this problem.


This paper will examine sensitivity to different modeling decisions of blockers. Specifically, in this paper, we will take into account the differences in three main dc-current blocking devices: line series capacitors, transformer neutral blockers, and substation ground approximation of neutral blockers. This will extend to a discussion of how different blocker placement representations vary results. Finally, we will contribute to the analysis of uniform and non-uniform electric field models and their impact on power loss output. 

This work relies on PowerModelsGMD.jl (abbrev. PMsGMD)\footnote{\url{https://github.com/lanl-ansi/PowerModelsGMD.jl}} for both dc network generation and analysis \cite{mate21-pmsgmd}. For the case of transformer neutral blockers, we relied on PowerWorld Simulator for results, as PMsGMD does not currently model these devices exactly and relies on the substation ground blocking approximation.


The general organization of the paper is as follows:
Section~\ref{dc-network-generation} discusses generation of dc networks.
Section~\ref{electric-fields} covers the calculations of the induced voltages on transmission lines.
Section~\ref{assumptions} describes the conditions that lead the assumption of implicit components.
Section~\ref{blockers} describes the modeling assumptions associated with different blocking options.
Section~\ref{implementation} describes the experimental procedure and case study system.
Finally, sections \ref{results} and \ref{conclusions} present results from the case study system and overall conclusions, respectively.

\section{DC Network Generation} \label{dc-network-generation}

\subsection{Constructing GMD Buses} \label{gmd-buses}

The main difference between the ac bus table and dc bus (GMD bus) table is the addition of substation grounding nodes, which do not have corresponding components in the ac model.
Since the additional branches in the dc network are the decomposition of transformer configurations, an extra bus at the ground of substations is required to represent all of the dc branches (GMD branches).

\begin{table}[!htbp] \scriptsize
    \caption{Mappings Between Transformer Terminals and Indices of Buses and Branches}
    \begin{center}
    {\renewcommand{\arraystretch}{1.1} 
    \begin{tabular}{cccc}
    \thicchline
    Terminal & ac & dc & dc \\
    & Bus & Bus & Branch \\
    \hline
    High-Side & $\eta(k)$ & $h(k)$ & $H(k)$ \\
    Low-Side & $\lambda(k)$ & $l(k)$ & $L(k)$ \\
    Tertiary & $\tau(k)$ & $t(k)$ & $T(k)$ \\
    Series & & $s(k)$ & $S(k)$ \\
    Common & & $c(k)$ & $C(k)$ \\
    Ground & & $g(k)$ & \\
    \thicchline
    \end{tabular}}
    \end{center}
    \label{table:transformer-index-mapping}
\end{table}
    
\subsection{Constructing GMD Branches} \label{gmd_branches}

GMD branches represent any path that GICs flow through. There are two main types of GMD branches: (1) the resistive components of ac transmission lines; and (2) windings of transformers that are connected to the substation ground. Additionally, implicit GMD branches are also generated from bypassed series capacitors as discussed in section \ref{series_capacitors}. 

\subsubsection{Lines}

The resistance of transmission lines in Ohms is calculated as follows:

\begin{equation}
R_{si} = R_{pu}^1\left(\frac{kV_{ll}^2}{3\text{\textit{MVA}}_{3\phi}}\right)
\end{equation}

\noindent In the above $R_{si}$ is the common-mode dc resistance of the branch, $R_{pu}^1$ is the positive-sequence resistance of the branch in per-unit, $kV_{ll}$ is the rated line-line voltage in $kV$, and $\textit{MVA}_{3\phi}$ is the three-phase base power in MVA.

\subsubsection{Transformers} \label{transformer_definitions} 

\begin{figure}[!htbp]
\centering
\includegraphics[width=0.25\textwidth]{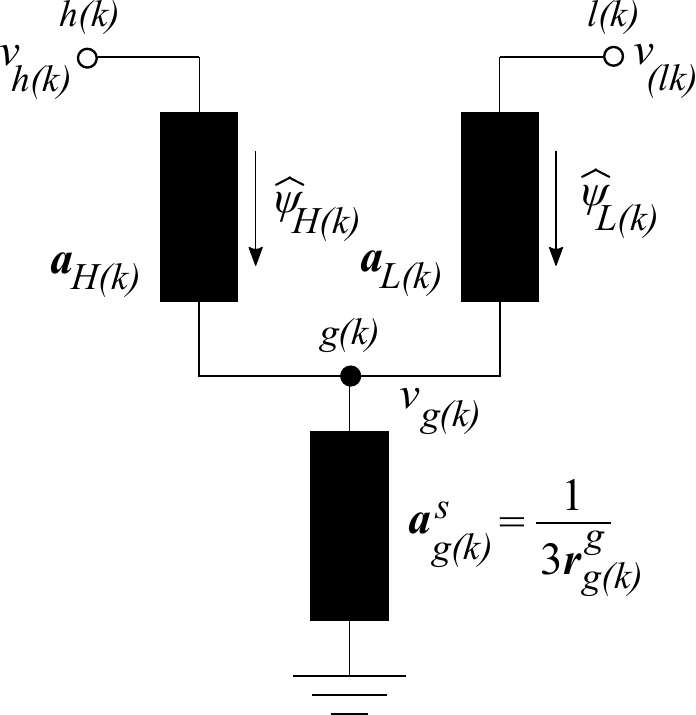}
\caption{DC equivalent circuit of a $Y_g$-$Y_g$ transformer  \cite{mate21-pmsgmd}}
\label{fig:gwye-gwye-equiv-circuit}
\end{figure}

\emph{$Y_g$-$Y_g$:} Both the winding configurations of the $Y_g$-$Y_g$ transformer have a connection to the ground of the substation. The dc current in these transformers can flow through the windings to ground, which completes a loop for GIC flow. To represent these extra paths for dc current, GMD branches are added on viable transformer windings. Particularly, any winding that is connected to ground has a GMD branch associated with it. The resistances of these branches are given as a third of the real part of the impedance of the windings to account for parallel phases. In addition, a GMD bus is added at the substation grounding node, as discussed in Section \ref{gmd-buses}. This is the substation grounding node, labeled $\textsl{g}(k)$, in Fig. \ref{fig:gwye-gwye-equiv-circuit} \cite{boteler_modeling_2016,zheng_effects_2014}.

\vspace{0.1in}

\emph{$\Delta$-$Y_g$:} These transformers do not have GMD branches from both buses to ground. Since the $\Delta$ windings do not have a direct connection to two GMD buses, GICs can not flow through them. Therefore, there is no GMD branch associated with the $\Delta$ windings. However, the grounded-wye ($Y_g$) side is represented the same way as the previous model. This is similar to Fig. \ref{fig:gwye-gwye-equiv-circuit}, but there is no connection from $h(k)$ to $\textsl{g}(k)$.

\begin{figure}[!htbp]
\centering
\includegraphics[width=0.15\textwidth]{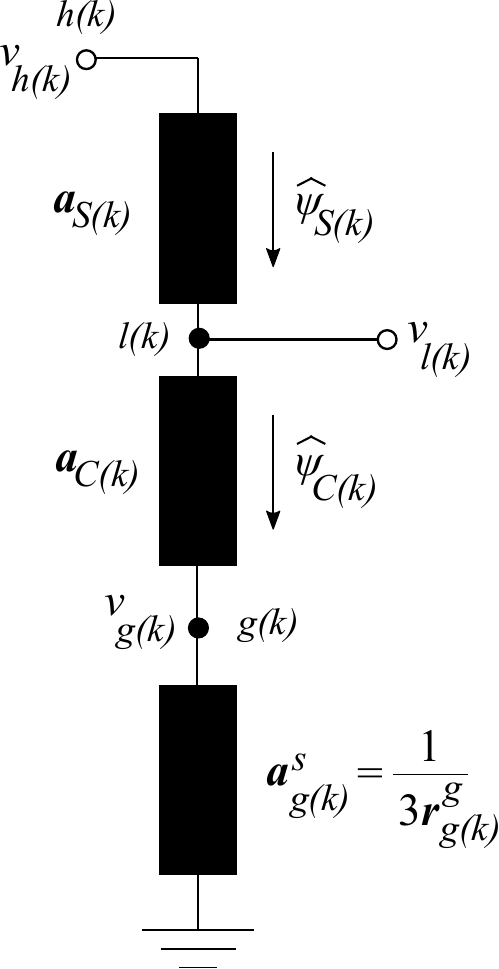}
\caption{DC equivalent circuit of a $Y_g$-$Y_g$ autotransformer  \cite{mate21-pmsgmd}}
\label{fig:gwye-gwye-auto-equiv-circuit}
\end{figure}

\emph{Autotransformers:} There are two GMD branches that we represent for each autotransformer, which are seen in Fig. \ref{fig:gwye-gwye-auto-equiv-circuit}. The series branch is from the $h(k)$ bus to the $l(k)$ bus. Since this part of the winding is connected to two different buses in the ac model, dc current can flow through it. In addition, similar to the $Y_g$ configurations, if the autotransformer is connected to the substation ground, the ``common" winding will also be represented as a GMD branch. This branch is justified once again because it connects two different GMD buses: $\textsl{g}(k)$ and $l(k)$ bus \cite{boteler_modeling_2016, zheng_effects_2014}.

\vspace{0.1in}

\begin{figure}[!htbp]
\centering
\includegraphics[width=0.25\textwidth]{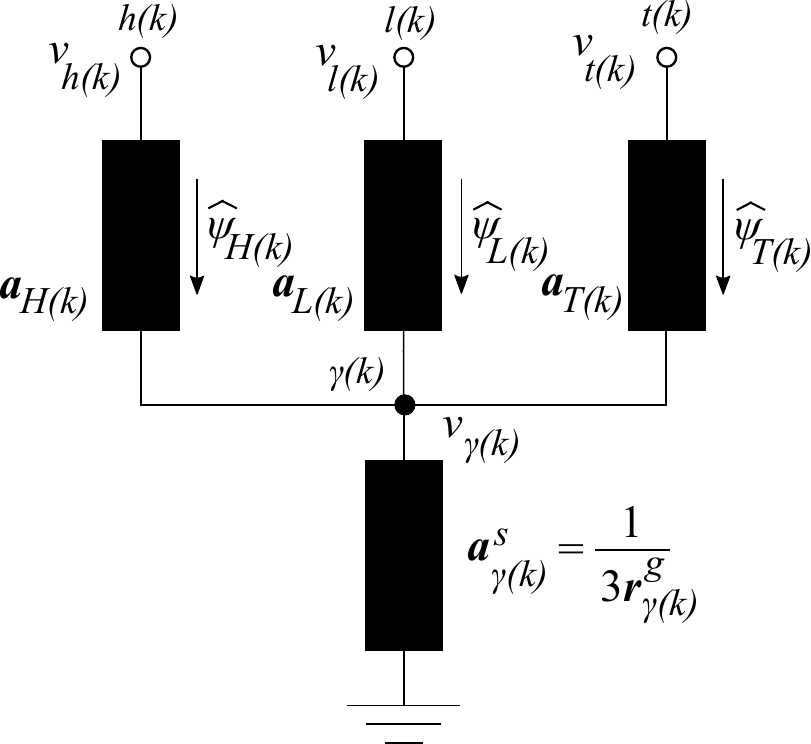}
\caption{DC equivalent circuit of a three-winding transformer}
\label{fig:3w-equiv-circuit}
\end{figure}

\emph{Three-Winding Transformers:} The primary difference between a three-winding transformer and the previously discussed transformers is the addition of a tertiary bus. In order to represent the three-winding transformer, the ac model contains three separate two winding transformers. The connecting node of these transformers is referred to as the ``star" bus. The windings of the star bus are not represented as GMD branches in the dc system; the star bus does not connect to any external gmd bus. Instead, $h(k)$, $l(k)$, $t(k)$ are connected to $\textsl{g}(k)$ independently through winding resistances. If any of the windings is not a $Y_g$ configuration, it is disconnected from the substation ground. This is shown in Fig. \ref{fig:3w-equiv-circuit} below \cite{boteler_modeling_2016,zheng_effects_2014}.

\vspace{0.1in}

\emph{Three-Winding Autotransformer:} If the three-winding transformer is an autotransformer between the $h(k)$ and $l(k)$ bus, the model is slightly changed. The two transformers connected to these buses in the ac system are modeled as separate autotransformers, such as Fig. \ref{fig:gwye-gwye-auto-equiv-circuit}. In this case, the star bus, $\sigma(k)$, can not be ignored, as it has a dc connection to both $h(k)$ and $l(k)$. This can be seen in the following Fig. \ref{fig:3w-auto-equiv-circuit}.

\vspace{0.1in}
    \begin{figure}[!htbp]
    \centering
    \includegraphics[width=0.25\textwidth]{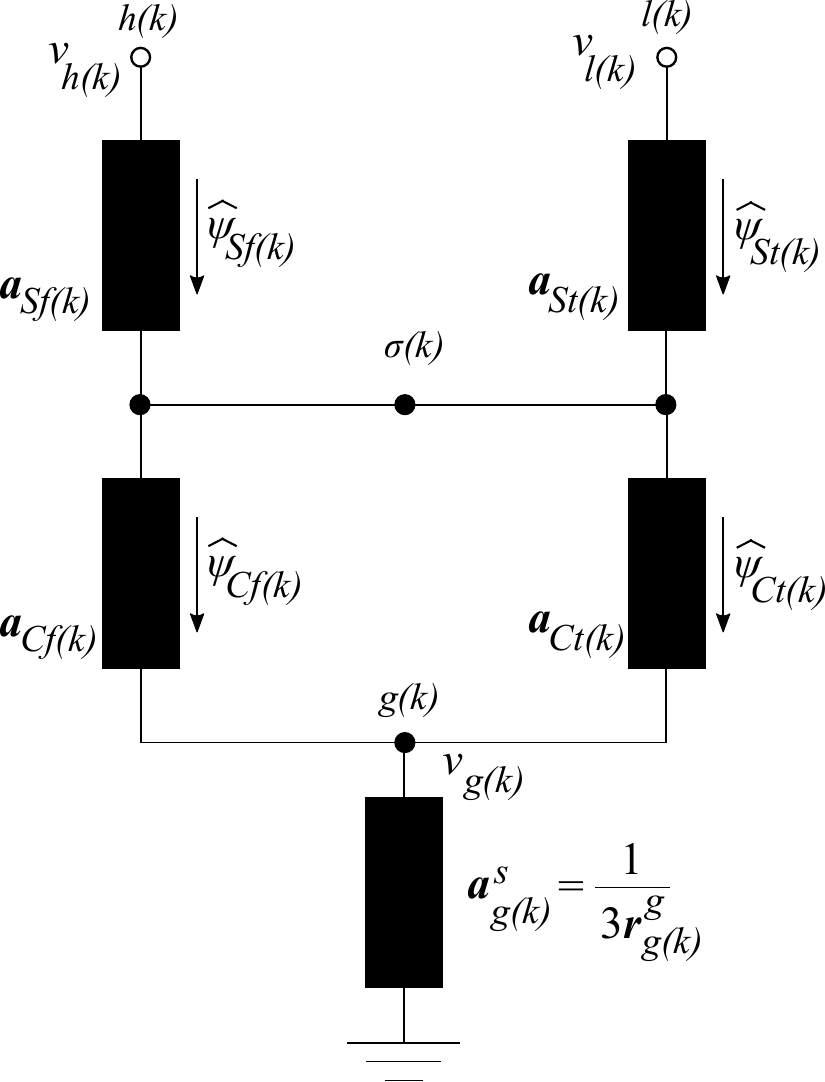}
    \caption{DC equivalent circuit of a $Y_g$-$a$-$\Delta$ transformer}
    \label{fig:3w-auto-equiv-circuit}
\end{figure}

The value of $a_{St(k)}$ is 1 $\mu\Omega$ in order to pull the voltage value of $l(k)$ onto $\sigma(k)$. In addition, the value of $a_{Cf(k)}$ is set to 1 M$\Omega$ so that the equivalent resistance from the $\sigma(k)$ to $g(k)$ is only that of the common winding of the overall transformer. Using these values, we end up with a dc representation of a single autotransformer, such that the series and common winding resistances are that of the real world circuit.

The relationship between ac nodes, dc nodes, and dc branches of transformers is summarized in Table \ref{table:transformer-index-mapping}.
        
\subsubsection{Power Loss Calculations} \label{ac-linking-constraints}

The first step for calculating the reactive power loss ($Q_{loss}$) in the transformers is to find the dc current magnitudes in each GMD branch. This is simply the result of Ohm's law $I_{dc} = g V_{br}$, where $g$ is conductance.

The next part of this process is to find the effective current in ac branches. For transformers, this value is utilized to find the power loss based on the given K-factor, where the K-factor is the constant of proportionality between reactive power consumption for a transformer and the weighted sum of GICs flowing though its terminals \cite{hutchins_effect_2011}. The linking here is done by referring to the dc currents in the mapped high side, low side, and tertiary GMD branches \cite{barnes2024review}.

The final step in calculating power loss is described by the following equation \cite{barnes2024review, overbye2012integration, hutchins_power_2016}:

\begin{equation}
    Q_{loss} = K I_{gic}  \left|\mathbf{V}_{\eta(k)}\right| I_{base}
\end{equation}

\noindent where $K$ is the K-factor of the transformer, $I_{gic}$ is the effective current in the ac transformer branch, $\left|\mathbf{V}_{\eta(k)}\right|$ is the voltage of the high-side bus, and $I_{base}$ is the base current at the high side of the transformer. This value results in the estimated power loss due to magnetic saturation in the transformer.

\section{Coupled Electric Field Calculations} \label{electric-fields}

The induced voltages across the transmission lines can be entered in two different ways: a CSV (\texttt{.csv}) line information file and a static electric field vector.

\subsection{CSV Line Information} \label{csv_lines}
The line information CSV is an optional input file, which contains the induced dc voltages across each ac line. These are then mapped to the corresponding GMD branch. The \texttt{GICInducedDCVolt} field in the CSV file is then copied into the \texttt{br\_v} parameter of the GMD branches. These value are passed to a matrix or optimization solver to arrive at the $Q_{loss}$ found in Section \ref{ac-linking-constraints}.

\subsection{Static Electric Field Vector}
Currently, PMsGMD calculates coupled dc line voltages based on time and space invariant electric field vectors. The angle $\theta$ of this vector is defined as $0^\circ$ pointing North and increases clockwise around the compass rose \cite{powerworld-defaults}. The magnitude is in units of V/km.

The latitude and longitude positions of substations are given in the GIC (\texttt{.gic}) file. The positions of other buses are mapped from the bus table in the GIC (\texttt{.gic}) file.
Given two sets of coordinates $(x_1, y_1)$ and $(x_2,y_2)$ that are represented in terms of (longitude, latitude), the following equations provide the displacement $\mathbf{d} = (d_e, d_n)$ between the points \cite{gic-application-guide}:

\begin{subequations}
\begin{align}
    \phi &= \dfrac{y_1 + y_2}{2} \\
    d_n &= M(\phi) \Delta y \label{length_north} \\
    d_e &= N(\phi) \Delta x \label{length_east} 
\end{align}
\end{subequations}

In the above, $\phi$ is the location of the latitude midpoint, $\Delta x = x_2 - x_1$ and $\Delta y = y_2 - y_1$. 
Equations \eqref{length_north} and \eqref{length_east} provide the northward and eastward distances given $M(\phi)$ and $N(\phi)$, which are the radius of curvature in the meridian plane and plane parallel to the latitude respectively. For this study we use the approximation below, which ensures that the sum of displacements along a closed path will sum to zero:

\begin{subequations} \label{approximations}
\begin{align}
    M(\phi) &= 110.574 \\
    N(\phi) &= 113.320 \cos\overline{\phi}
\end{align}
\end{subequations}

\noindent In the above $\overline{\phi}$ is the arithmetic mean of the latitude midpoints for the network under study. These values are used along with the electric field ($\vec{E}$) to gain voltage difference between two points. 

\begin{equation}
    V_{br} = \vec{E}\cdot\vec{d} = \left|\Vec{E}\right| \left(d_n \cos{\theta} + d_e \sin{\theta}\right) 
\end{equation}

This formulation is utilized for each GMD branch by inputting the branch positions for points $(x_1, y_1)$ and $(x_2,y_2)$ above. These expressions assume a constant electric field. In order to input data for a varying field, refer to Section \ref{csv_lines}.

\section{Implicit Components} \label{assumptions}

\subsection{Implicit Generator Step-up Transformers} \label{generator-implicit}

If a generator is connected to a bus that has a nominal voltage $\geq$ 30 kV and the generator status is 1, then an implicit generator step-up (GSU) is added to the bus. The configuration of this transformer is $\Delta$-$Y_g$. A dc branch is inserted on the $Y_g$ side from the generator bus to the substation. The resistance of this branch is determined by the generator base nominal voltage through Table \ref{table:gsu-impedance-lookup}, below \cite{birchfield2016statistical}.

\begin{table}[!htbp] \scriptsize
\caption{Implicit GSU Resistance vs. Nominal Voltage}
\begin{center}
{\renewcommand{\arraystretch}{1.1} 
\begin{tabular}{cc}
\thicchline
Generator Base (kV) & Winding Resistance ($\mu\Omega$)\\
\hline
765.0 & 1.089\\
500.0 & 1.667\\
345.0 & 2.416\\
230.0 & 3.623\\
161.0 & 5.176\\
138.0 & 6.039\\
115.0 & 7.246\\
\thicchline
\end{tabular}}
\end{center}
\label{table:gsu-impedance-lookup}
\end{table}

The extra branch is only added to the dc network. However, since this transformer doesn't exist in the ac model, it is not added to the $Q_{loss}$ table. The only affect from the assumed GSU is on the $Q_{loss}$ of the other transformers.

\subsection{Implicit GMD Branches}

    Implicit dc branches are placed between every bus and their substation with 25 k$\Omega$ resistance. PowerWorld Simulator likely adds this branch to help their matrix solver. Without any grounding points, the inverse of a conductance matrix ($G$-matrix) results in a singularity. This is because if the diagonal values are equal to the negative sum of their corresponding off-diagonal values, then the determinant of the matrix is 0. Thus, by adding grounding to the buses, we create a stronger diagonal, which prevents singularities in smaller sections of the matrix.

\subsection{Assumed Series Capacitors} \label{series_capacitors}

The representation of series capacitors has three modes in PowerWorld Simulator: \texttt{OPEN}, \texttt{CLOSED}, and \texttt{BYPASSED}. When the capacitor is \texttt{BYPASSED}, there is a branch with negligible resistance placed in parallel with the capacitor. This is represented as a small 5 m$\Omega$ resistor in order to prevent an infinite entry in the conductance matrix. An ac branch is assumed to be a \texttt{BYPASSED} capacitor if the given resistance is 0 $\Omega$. In the dc network, an open branch with a series capacitor is not converted to a GMD branch, as no dc current can flow through it. 

\section{Blocker Modeling} \label{blockers}

There are several different mitigation techniques for the aforementioned GICs. To solve placement optimization problems, PMsGMD allows for the placement of blockers at different points in the circuit. In the following sections, we will discuss the different methods and the varying effects with each.

\subsection{GIC Blocker Modeling} \label{gic-blockers}

One of the methods of reducing the reactive power loss is by placing a GIC blocker at different points between the transformer neutral point to ground. There are two primary options for this: neutral blocking and substation blocking.

\emph{Neutral Blocking:} Placing a blocker at the neutral point of the transformer disconnects the neutral node from $\textsl{g}(k)$. In the basic case of the $Y_g$-$Y_g$ transformer, shown in Fig. \ref{fig:gwye-gwye-equiv-circuit}, the neutral blocker is placed between a transformer neutral node and ground node $\textsl{g}(k)$. This means that no GICs can flow through the transformer. However, for an autotransformer, such a placement still allows for the flow of current from the dc high-side node $h(k)$ to the dc low-side $l(k)$ nodes in Fig. \ref{fig:gwye-gwye-auto-equiv-circuit}.

\emph{Substation Blocking:} An approximation for the neutral blocker is to place a single blocker at the ground of the substation. This blocker would replace the $a^s_{\textsl{g}(k)} $ resistor in Fig. \ref{fig:gwye-gwye-equiv-circuit}. The important distinction here is that the neutral points for multiple transformers will still be connected if they are at the same substation. Therefore, in the case of circulating current from nonuniform fields (discussed further in Section \ref{uniform-field}), there will still be $Q_{loss}$.

\subsection{Series Capacitor Blocking}

As discussed in section \ref{series_capacitors}, adding a series capacitor to a transmission line means that there will be no equivalent GMD branch generated. This is because the impedance for a dc current through a capacitor is infinite. Therefore, if a series capacitor is placed on every line, we expect to see no power losses in the whole system. The problem with this method is that an increased reactance will result in a phase shift along ac branches.

\subsection{Uniform Electric Fields Assumption} \label{uniform-field}
In the case of uniform electric fields, there is no circulating current, which is defined as current passing around loops in the network that do not contain ground. This is known because a line integral around a closed path of a conservative field evaluates to 0 V. Therefore, there is no net voltage to allow for circulating current in the transmission lines.

Transformer neutral blocking would be enough if there was no circulating current. However, by accounting for variations of the induced electric field during a GMD, circulating current can flow through loops of autotransformers. This is because the neutral blocking does not prevent GICs in the series branch ($\hat{\Psi}$ in Fig. \ref{fig:gwye-gwye-auto-equiv-circuit}) from flowing.

A uniform electric field is conservative, such that the integral of the field along a closed contour is zero \cite{lorrain1970electromagnetic}:

\begin{equation} \label{faraday}
    \oint \mathbf{E} \cdot \mathbf{}{dl} = \sum_{k \in \mathcal{L}} V_k =  0
\end{equation}

\noindent Therefore, by solving the matrix form of the dc circuit with mesh current analysis we get the following:

\begin{subequations}
\begin{align}
    \mathbf{R} \Vec{I} = \vec{V} = \vec{0}\\
    \vec{I} = \mathbf{R}^{-1} \vec{V} = \vec{0}
\end{align}
\end{subequations}

\noindent This implies that given a uniform field and a fully determined system, we have no circulating current in our network; circulating currents only occur in loops, and every loop consists of resistances with 0 V induced. This result does not depend on whether the graph representation of the circuit is planar. It only relies on the circuit containing loops in a uniform field. 

\section{Implementation and Case Study} \label{implementation}

\subsection{Case Study System}

\vspace{0.1in}
\begin{figure}[!htbp]
\centering
\includegraphics[width=0.47\textwidth]{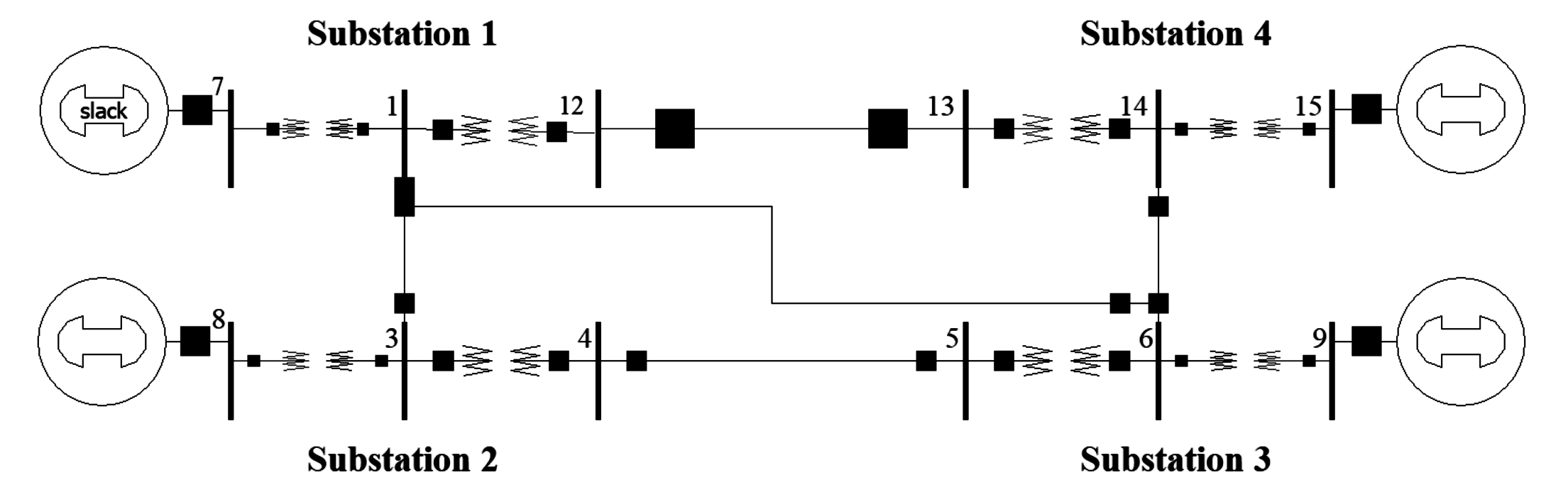}
\caption{GIC blocker modeling test case}
\label{fig:test-case}
\end{figure}

This circuit was specifically designed to demonstrate the differences between the different blocker placements. There are two main meshes here, which are affected differently by where the blocking device is located. The loop with only autotransformers will have a closed path for GICs despite having a neutral blocker. This distinguishes the results from the series capacitor case, which opens all loops.
This case is available online\footnote{\url{https://github.com/lanl-ansi/PowerModelsGMDLib}} for the sake of reproducibility.

There are four substations with three buses each in this system. Substations 1, 3, and 4 span Canada and Alaska from East to West, and substation 2 is placed in Houston. A GSU is placed at each generator in addition to the other transformers described earlier. The “NERC TPL-007.1 and TPL-007.2.” is used for the scaling function for the electric field, which has a magnitude of 1 V/km with a $90^\circ$ bearing from north.

\subsection{Procedure for Numerical Experiments}

When running the case from PowerWorld through PMsGMD for dc network generation and power loss analysis, the first step is to extract the required input files. We utilized the \texttt{gmdtool} command-line program\footnote{\url{https://github.com/bluejuniper/gmd-tools}} to convert the \texttt{.pwb} PowerWorld case into either the extended MatPower format \cite{mate21-pmsgmd} or (\texttt{.raw}, \texttt{.gic}, \texttt{.csv}) format employed in this paper. The \texttt{gmdtool} executable interfaces with PowerWorld Simulator to export case data needed for GIC analysis problems.

The PMsGMD software now includes functionality for dc network generation which previously relied on \texttt{gmd-tools}. Additionally, while PMsGMD is designed primarily for power systems optimization problems involving GMD events, it also includes the linear matrix solver for verification that is also used for this study \cite{Lehtinen1985}. PMsGMD is built on top of PowerModels, which makes use of the JuMP optimization solver interface. This enables a fast construction of networks from input files because Julia is a just in time (JIT) language. 

The experiment is run in two stages: on a uniform electric field and a non-uniform electric field. For both these scenarios, power loss is calculated for each transformer for four blocker placement locations: no blocker, substation blocker, neutral point blocker, and series capacitor. Each blocker type is placed at 100\% of possible locations discussed in Section \ref{blockers}. The losses at each transformer are then recorded for the eight total scenarios.

Finally, for the results found in Section \ref{results}, we utilized PowerWorld's solver with neutral blockers, as this is not something yet implemented in PMsGMD. Additionally, for the electric field calculations, PowerWorld generated the line voltages we inputted into PMsGMD for the non-uniform case. Since the electric field calculations are based on approximations, the net displacement around a loop does not end up to exactly 0 km. In order to force this property to be true, Fig. \ref{fig:uniform-results} utilizes \eqref{approximations} so that \eqref{faraday} is exactly 0 V.

\section{Results} \label{results}

\vspace{0.1in}
\begin{figure}[!htbp]
\centering
\includegraphics[width=0.4\textwidth]{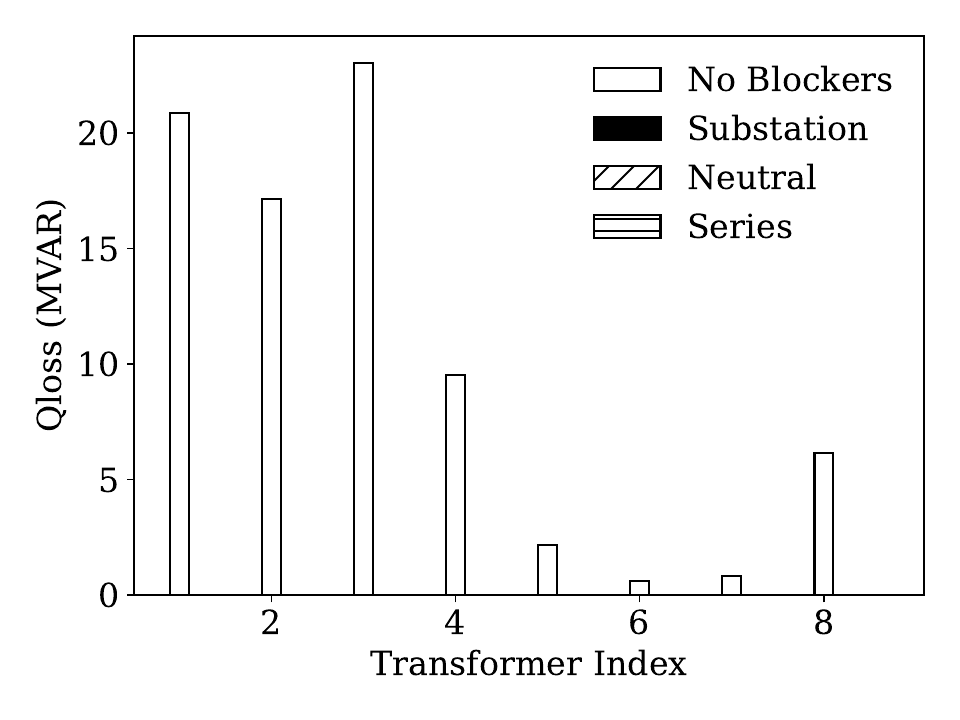}
\caption{Comparison of $q_{loss}$ with different blocker placements in a uniform electric field}
\label{fig:uniform-results}
\end{figure}

\vspace{0.1in}
\begin{figure}[!htbp]
\centering
\includegraphics[width=0.4\textwidth]{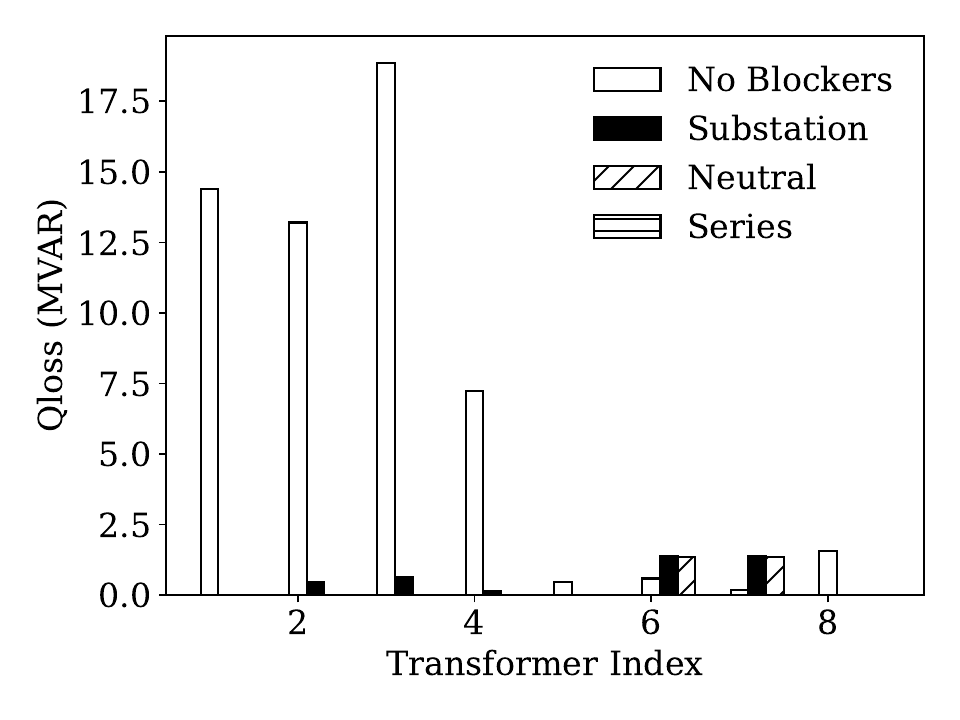}
\caption{Comparison of $q_{loss}$ with different blocker placements in a non-uniform electric field}
\label{fig:non-uniform-results}
\end{figure}

Fig. \ref{fig:uniform-results} shows the power loss on each transformer with a uniform representation of the electric field. With this approximation, it is clear that all blocker types result in 0 MVAR power loss with 100\% placement. Therefore, in the case where a uniform electric field is applicable, the blocker representation doesn't have an effect on the result due to \eqref{faraday}.

Fig. \ref{fig:non-uniform-results} illustrates the power loss for each transformer in a non-uniform electric field. In this case, due to closed loops having induced voltages across them, the substation blocker approximation of the neutral blocker seems to make a noticeable difference. Substation blocking overestimates the power loss due to additional meshes in the dc network.

\section{Conclusions} \label{conclusions}

Modeling GICs allows for flexibility in many ways. Though the algorithms for generating the dc network are generally consistent, the interpretation of different parameters and choices of device placement allows for a variety of different results. 
One choice is the representation of the blocking devices. The results in Section \ref{results} demonstrate the sensitivity of the outcome based on the blocking device within the context of different electric fields. The substation approximation of a transformer neutral blocker is demonstrated to be reasonable, introducing only minor error in non-uniform electric fields. 

In the case of a uniform electric field, 100\% blocker placement will always drive the power loss to 0 MVAR regardless of blocker type. However, in non-uniform electric fields, this conclusion is not necessarily true and series blocking devices are needed to drive power loss to zero. This result motivates caution in analysis of GICs, as it has been common practice to use uniform electric fields, which could potentially obscure system vulnerabilities. 

The future use of this modeling formulation could extend to the analysis of other effects of GICs. Due to the magnetic saturation of the transformers, the output waveform of the transformer contains clipping of a pure sinusoid. This indicates that control devices may trip, such as static VAR compensators (SVCs) during the March 1989 solar storm, in the rest of the system due to induced harmonics \cite{boteler201921st}. Such an outcome also motivates tying the modeling described in the paper back to an ac power flow analysis. The impact of a solar flare has the capability of causing various disturbances in the energy infrastructure, but by considering the details of GIC modeling, we position ourselves to better identify such threats before they occur.

\section*{Acknowledgment}
We thank Russell Bent (Applied Mathematics and Plasma Physics Group at Los Alamos National Laboratory, Los Alamos, NM 87545 USA) for assistance with with conceptual aspects of this paper.

\bibliographystyle{unsrt}
\bibliography{refs}
\end{document}